# Compatibility of symmetric quantization with general covariance in the Dirac equation and spin connections

A. D. Alhaidari [a,†] and A. Jellal [a,b]

[a] *Saudi Center for Theoretical Physics, P. O. Box 32741, Jeddah 21438, Saudi Arabia*
[b] *Theoretical Physics Group, Faculty of Sciences, Chouaib Doukkali University, 24000 El Jadida, Morocco*

**Abstract**: By requiring unambiguous symmetric quantization leading to the Dirac equation in a curved space, we obtain a special representation of the spin connections in terms of the Dirac gamma matrices and their space-time derivatives. We also require that squaring the equation give the Klein-Gordon equation in a curved space in its canonical from (without spinor components coupling and with no first order derivatives). These requirements result in matrix operator algebra for the Dirac gamma matrices that involves a universal curvature constant. We obtain exact solutions of the Dirac and Klein-Gordon equations in 1+1 space-time for a given static metric.

## 1. INTRODUCTION

Understanding the connection between quantum theory (mechanics and fields) and gravity continues to be one of the main tasks in contemporary physics that proved to be highly nontrivial and very demanding. Formulation of quantum gravity is still far from being successful or even satisfactory. A consistent unification of quantum theory and gravity must first address the state of a single elementary particle in a gravitational background. Consequently, sustained efforts have been applied to find a systematic and appropriate formulation of the relativistic equation of motion for the lowest spin particles (spin-0 and spin-$\frac{1}{2}$) in a curved space-time. That is, the extension of the Klein-Gordon and Dirac equations from flat space to a curved space. One of the interesting problems in this connection is the extent to which spin has an effect on the quantum gravitational phenomena. For example, it has been shown in [1] that the spectrum of spin-0 and spin-$\frac{1}{2}$ particles in a constant gravitational field differ by an amount of $\sqrt{mg\hbar c}$, where *m* is the rest mass of the particle and *g* is the acceleration of gravity. Although weak, this is a significant difference that shows the influence of spin in gravitational interaction. Moreover, unambiguous observation of the influence of gravity on the behavior of fermions is one of the major motivations to study the Dirac equation in curved space. An example is the quantum effects on neutrons in a classical gravitational field [2-6]. In recent years, investigation of Dirac particles in virtual gravitational fields have been at the center of interest in condensed matter physics in the context of studies of the amazing properties of graphene [7-10]. It was shown that it is possible to simulate some of these properties by coupling the Dirac fermions to an "artificial" gravitational field; specifically, to consider the physics of massless Dirac particles in a 2+1 curved space-time. These results exhibit rare and direct connection between gravity and quantum mechanics and constitute another strong motivation to the study of the Dirac equation in curved space.

---

[†] Present address: Al-Anwar Tower #4202, Al-Qasba, Sharjah, United Arab Emirates
Email: haidari@sctp.org.sa

The greatest difficulties in these studies arise from the covariant generalization of the Dirac equation [11,12] and its uniqueness. Due to the complexity of the Dirac equation (a system of coupled partial differential equations), the number of exact solutions even in the special theory of relativity remained very limited. There are two types of difficulties that occur in the solution of the Dirac equation in special relativity. The first is due to the physical nature of the problem; in particular, the geometry of the external field. The second is purely mathematical and is related to the choice of coordinates. On the other hand, a complete theory of separation of variables for the Dirac equation in a curved space-time has yet to be developed. Nonetheless, it is common knowledge that separation of variables in the Dirac equation is easier for the massless case and in the context of the Kerr geometry [13-15]. The connection between separation of variables and matrix first-order differential operators commuting with the Dirac Hamiltonian has traditionally been the prime focus in such developments. However, in [16] the separation problem was solved provided that the squared Dirac equation (or the Klein-Gordon equation) is reduced to two independent differential equations of second order (i.e., it admits diagonalization).

The equation of relativistic quantum mechanics was formulated in the early part of last century by Paul Dirac [17]. It describes the state of electrons in a way consistent with quantum mechanics and special relativity. The physics and mathematics of the Dirac equation is very rich, illuminating and provides a theoretical framework for different physical phenomena that are not present in the nonrelativistic regime such as the Klein paradox, super-criticality [17-19] and the anomalous quantum Hall effect in graphene [20,21]. The free Dirac equation in its classical representation is the square root of Einstein's relativistic statement $p^2 = m^2c^2$, where $p$ is the space-time linear momentum vector. It is written as $\gamma^\mu p_\mu = mc$, where $\{\gamma^\mu\}_{\mu=0}^{n}$ is a set of square matrices that are related to the metric tensor of the $n+1$ space-time by $\{\gamma^\mu, \gamma^\nu\} = 2g^{\mu\nu}\mathbf{1}$ and repeated indices are summed over. In flat space, the metric is constant and, thus, the Dirac gamma matrices $\{\gamma^\mu\}$ are independent of space and time. Therefore, with $p_\mu \to i\hbar\partial_\mu$, quantization is straightforward and the Dirac equation for a free spinor is written as $i\hbar\gamma^\mu\partial_\mu\psi = mc\psi$, where $\psi$ is the multi-component wavefunction. However, in a curved space, where the metric is not constant, these matrices are space-time dependent. Thus, quantization of the classical term $\gamma^\mu(x)p_\mu$ becomes a nontrivial issue that may involve ordering ambiguity. However, it is known that symmetric quantization of the classical phase space function product $f(x)g(p)$ is not ambiguous if and only if $f(x)$ or $g(p)$ is linear; which is the case here. Specifically, symmetric quantization of the classical phase space product $f(x)p$ is $\frac{1}{2}\left(f^\alpha p f^\beta + f^\beta p f^\alpha\right)$, where $\alpha$ and $\beta$ are arbitrary real parameters such that $\alpha + \beta = 1$. In configuration space where $p \to i\hbar\frac{d}{dx}$, symmetric quantization gives

$$\tfrac{1}{2}\left(f^{\alpha}pf^{\beta}+f^{\beta}pf^{\alpha}\right)=i\hbar f(x)\tfrac{d}{dx}+\tfrac{i}{2}\hbar df/dx ,\tag{1}$$

which is independent of the choice of parameters. Therefore, symmetric quantization of $\gamma^{\mu}(x)p_{\mu}$ gives $i\hbar\gamma^{\mu}(x)\partial_{\mu}+\tfrac{i}{2}\hbar\left(\partial_{\mu}\gamma^{\mu}\right)$. On the other hand, covariant generalization of the Dirac equation in a curved space is achieved by introducing the covariant derivative via the substitution $\partial_{\mu}\to\partial_{\mu}+\Gamma_{\mu}$, where $\{\Gamma_{\mu}\}$ are the $n+1$ spin connection matrices. Thus, $\gamma^{\mu}(x)p_{\mu}\to i\hbar\gamma^{\mu}(x)\partial_{\mu}+i\hbar\gamma^{\mu}\Gamma_{\mu}$. Therefore, nominal compatibility of symmetric quantization with general covariance gives a special representation of the contracted spin connections $\gamma^{\mu}\Gamma_{\mu}$ in terms of the space-time divergence of the gamma matrices. More precisely,

$$\gamma^{\mu}\Gamma_{\mu}=\tfrac{1}{2}\partial_{\mu}\gamma^{\mu}.\tag{2}$$

The covariant generalization of the Dirac equation to curved space was independently developed long ago by Weyl [22] and by Fock [23], which is known in the literature as Dirac-Fock-Weyl (DFW) equation. Recently, two alternative versions of the Dirac equation in a curved space-time were proposed in [24]. These obey the equivalence principle in a direct and explicit sense, whereas the DFW equation obeys the same only in an extended sense. The present work, which is complementary to those cited above, may constitute a measurable contribution in the pursuit of a systematic formulation of the Dirac and Klein-Gordon equations in a curved space. Specifically, we use the special representation of the spin connections obtained above to write the Dirac equation in curved space. We will also introduce a matrix operator algebra involving the Dirac gamma matrices such that the Klein-Gordon equation that results from squaring the Dirac equation is in its canonical form with no coupling among the spinor components and without first order derivatives. As a result, we find that arbitrary spin connections and/or vierbeins are not needed for writing down the Dirac equation in a curved space. It is true that this problem has long been treated in full generality with the use of vierbeins and spin connections that make clear how covariance under general coordinate and local Lorentz transformations is achieved. However, the prescription suggested here set aside vierbeins and spin connections in favor of a simple and consistent formulation of the Dirac equation in a curved space. Thus, in the language of vierbeins and spin connections, the present formulation leads to the Dirac equation in a suitable gauge (e.g. in a given choice of tangent frame). The option of not using vierbeins has appeared in the earlier literature [25-29] though often without any proof that spin connections exist.

We conclude this work with an example where we choose a static metric in 1+1 space-time and obtain exact solutions for free spin-0 and spin-$\tfrac{1}{2}$ relativistic particles in this gravitational background. In the following section, we start by defining the matrix operator algebra and point out its correspondence with the quantum mechanical algebra and the classical Poisson bracket algebra.

## 2. OPERATOR ALGEBRA FOR THE DIRAC GAMMA MATRICES: DIRAC & KLEIN-GORDON EQUATIONS

The covariant generalization of the free Dirac equation ($i\gamma^\mu \partial_\mu \psi = m\psi$) in a curved space-time of dimension $n+1$ reads as follows is

$$i\gamma^\mu \left(\partial_\mu + \Gamma_\mu\right)\psi = m\psi, \qquad (3)$$

where we have adopted the conventional relativistic units $\hbar = c = 1$. Now, we propose the following one-parameter Dirac equation in a curved space

$$i\left(\gamma^\mu \partial_\mu - \lambda\Omega\right)\psi = m\psi, \qquad (4)$$

where $\lambda$ is a dimensionless parameter and for $\lambda = -1$, $\Omega = \gamma^\mu \Gamma_\mu$. The transformation properties of the space-time dependent matrix $\Omega$ is the same as that of $\gamma^\mu \Gamma_\mu$ and results from the covariance of Eq. (4) under general coordinate transformation and local spinor transformations. If we adopt the representation of the spin connections given by Eq. (2), then $\Omega = \frac{1}{2}\partial_\mu \gamma^\mu$. Iteration of Eq. (4) (i.e., squaring the equation) should result in the Klein-Gordon equation which reads as follows

$$\left[-\gamma^\mu \gamma^\nu \partial_\mu \partial_\nu + \left(-\slashed{\partial}\gamma^\nu + \lambda\{\Omega,\gamma^\nu\}\right)\partial_\nu + \lambda\slashed{\partial}\Omega - \lambda^2 \Omega^2\right]\psi = m^2\psi, \qquad (5)$$

where $\slashed{\partial} = \gamma^\mu \partial_\mu$. However, the conventional Klein-Gordon equation in a curved space is normally written as

$$\left[\frac{1}{\sqrt{-g}}\partial_\mu\left(g^{\mu\nu}\sqrt{-g}\,\partial_\nu\right) + m^2\right]\psi = \left[g^{\mu\nu}\left(\partial_\mu\partial_\nu + \Gamma^\sigma_{\mu\nu}\partial_\sigma\right) + m^2\right]\psi = 0, \qquad (6)$$

where $g$ is the determinant of the metric tensor, $\Gamma^\sigma_{\mu\nu} = \frac{1}{2}g^{\sigma\rho}\left(\partial_\mu g_{\rho\nu} + \partial_\nu g_{\rho\mu} - \partial_\rho g_{\mu\nu}\right)$ and $\{g_{\mu\nu}\}$ are elements of the inverse of the metric tensor. In Eq. (6), we used the logarithmic derivative identity: $\frac{1}{\sqrt{-g}}\partial_\mu \sqrt{-g} = \Gamma^\sigma_{\mu\sigma}$. Compatibility of this version of the equation with Eq. (5) results in the following algebra for $\Omega$ and the gamma matrices

$$\slashed{\partial}\gamma^\mu = \lambda\{\Omega,\gamma^\mu\} + g^{\sigma\nu}\Gamma^\mu_{\sigma\nu}, \qquad (7a)$$

$$\slashed{\partial}\Omega = \tfrac{\lambda}{2}\{\Omega,\Omega\}. \qquad (7b)$$

On the other hand, if we require that the Klein-Gordon equation be in its canonical form where not only spin components coupling is absent but also no first order derivatives, then we obtain the following alternative algebra

$$\slashed{\partial}\gamma^\mu = \lambda\{\Omega,\gamma^\mu\}, \qquad (8a)$$

$$\slashed{\partial}\Omega = \tfrac{\lambda}{2}\{\Omega,\Omega\} + \lambda R\mathbb{I}, \qquad (8b)$$

where $R$ could be taken as the scalar curvature constant of space and time[‡] and $\mathbb{I}$ is a diagonal space-time dependent matrix. Because of this alternative algebra, Eq. (5) becomes

$$\left[g^{\mu\nu}\partial_\mu\partial_\nu - \lambda^2 R\mathbb{I}\right]\psi = -m^2\psi. \qquad (9)$$

This is the Klein-Gordon equation in a curved space in its simple canonical form (no first order derivatives and no spinor components coupling). Therefore, here we will implement the algebra given by Eq. (8) and adopt the resulting Klein-Gordon equation (9).

---

[‡] Alternatively, it could also be taken as the cosmological constant.

Now, the algebra (8) has correspondence with ordinary nonrelativistic quantum mechanics where, for example, $i\hbar \partial_t F = [H, F]$ and $-i\hbar \vec{\nabla} F = [\vec{p}, F]$ [30]. Thus, the commutator $[\,,\,] \to$ the anti-commutator $i\{,\}$ and $\hbar \to \lambda^{-1}$, whereas functions are replaced by matrices. However, in nonrelativistic quantum mechanics we also have $i\hbar \partial_t \psi = H\psi$ whereas here $\frac{1}{\lambda}\tilde{\partial}\psi = (\Omega - im/\lambda)\psi$, which maintains the correspondence only in the massless case ($m = 0$). Another correspondence could also be drawn with the Poisson bracket algebra of classical mechanics, where the Poisson bracket corresponds to the anti-commutator [31]. The spin connection and the Riemann-Christoffel connection $\Gamma^\nu_{\mu\sigma}$ are related as [32]

$$\partial_\mu \gamma^\nu + [\Gamma_\mu, \gamma^\nu] + \Gamma^\nu_{\mu\sigma} \gamma^\sigma = 0. \tag{10}$$

Compatibility of this relation with the matrix algebra (8) for $\lambda = -1$ gives a trivial representation of the spin connections as $\Gamma^\nu = \frac{1}{2} g^{\mu\sigma} \Gamma^\nu_{\mu\sigma}$, which is a set of functions rather than matrices. Thus, in the Dirac equation (4), the matrix nature of $\Omega$ is in the linear combination of the $\{\gamma^\mu\}$ not in $\Gamma_\mu$ itself. If we adopt the representation (2), then we may also write $\Gamma^\nu = \frac{1/2}{n+1} \text{Tr}(\gamma^\nu \partial_\mu \gamma^\mu)$. On the other hand, compatibility of relation (10) with the alternative algebra (7) for $\lambda = -1$ gives $\Gamma^\nu = g^{\mu\sigma} \Gamma^\nu_{\mu\sigma}$.

Therefore, Eq. (4) with the $n+2$ matrices $\{\Omega, \gamma^\mu\}_{\mu=0}^n$ satisfying the matrix algebra (8) is taken here as the Dirac equation in a curved space-time whose metric is $g^{\mu\nu}$ and scalar curvature constant is $R$. The corresponding Klein-Gordon equation for spin-0 particle is Eq. (9). Note that in the flat space limit the $\gamma$ matrices are constants with $R = 0$ and we recover trivially the traditional Dirac and Klein-Gordon equations of special relativity provided that $\lim_{R \to 0} g^{\mu\nu} = \delta^{\mu\nu}$, where $\delta^{\mu\nu}$ is the flat space-time metric. For a given space-time metric $g^{\mu\nu}$, it is a challenging task in representation theory to account for the $n+1$ dimensional space of square matrices $\{\gamma^\mu\}$ satisfying $\{\gamma^\mu, \gamma^\nu\} = 2g^{\mu\nu} \mathbf{1}$ and the algebra of Eqs. (8). In the following section, we give an example of our formulation in 1+1 space-time with a static metric.

### 3. AN EXAMPLE: 1+1 SPACE-TIME WITH STATIC METRIC

For $n = 1$, let us consider the quantum mechanical behavior of a free spinor particle influenced only by the gravitational field in 1+1 space-time with a static metric. In this case the Dirac gamma matrices are 2×2 time-independent matrices and $\Omega = \frac{1}{2}\gamma'_1$, where the prime stands for the derivative with respect to $x$. The resulting kinematic equations (the metric relations and matrix algebra) are:

$$\gamma_0^2 = g^{00} \begin{pmatrix} 1 & 0 \\ 0 & 1 \end{pmatrix}, \quad \gamma_1^2 = g^{11} \begin{pmatrix} 1 & 0 \\ 0 & 1 \end{pmatrix}, \tag{11}$$

$$\gamma_0\gamma_1 + \gamma_1\gamma_0 = 2g^{01}\begin{pmatrix}1 & 0\\ 0 & 1\end{pmatrix},\tag{12}$$

$$\tfrac{2}{\lambda}\gamma_1\gamma_0' = \gamma_1'\gamma_0 + \gamma_0\gamma_1',\tag{13}$$

$$\left(\tfrac{2}{\lambda}-1\right)\gamma_1\gamma_1' = \gamma_1'\gamma_1,\tag{14}$$

$$\tfrac{2}{\lambda}\gamma_1\gamma_1'' = (\gamma_1')^2 + 4R\begin{pmatrix}p & 0\\ 0 & q\end{pmatrix},\tag{15}$$

where we wrote $\mathbb{I} = \begin{pmatrix}p & 0\\ 0 & q\end{pmatrix}$ and $p$ and $q$ are two independent functions of $x$. Indices are placed lower on the gamma matrices only for improved presentation. Equation (11) shows that the most general form that the gamma matrices can take is $\gamma = \begin{pmatrix}a & b\\ c & -a\end{pmatrix}$, where $\{a,b,c\}$ are independent functions of $x$. However, Eq. (12) entail that both matrices cannot be diagonal at the same time otherwise the metric becomes singular. Now, Eq. (14) requires that $\lambda = 1$ and $\gamma^1 = \begin{pmatrix}a & \rho\\ \tau & -a\end{pmatrix}$ where $\rho$ and $\tau$ are arbitrary constant parameters. Substituting this in Eq. (15) dictates that $\rho = \tau = 0$ and gives

$$p(x) = q(x) = \tfrac{1}{4R}\left(2aa'' - a'^2\right).\tag{16}$$

For this representation of $\gamma^1$ we impose Eq. (13) resulting in the following most general pair of gamma matrices in 1+1 space-time with a static metric that satisfy the algebra (8) with $\Omega = \tfrac{1}{2}\partial_\mu\gamma^\mu$:

$$\gamma^1 = a\begin{pmatrix}1 & 0\\ 0 & -1\end{pmatrix},\quad \gamma^0 = \begin{pmatrix}\mu a & \alpha\\ \beta & -\mu a\end{pmatrix},\tag{17}$$

where $\alpha$, $\beta$ and $\mu$ are constants. Therefore, the associated space-time metric tensor has the following components

$$g^{11} = a^2,\tag{18a}$$

$$g^{01} = g^{10} = \mu a^2,\tag{18b}$$

$$g^{00} = \alpha\beta + \mu^2 a^2.\tag{18c}$$

In the flat space limit ($R \to 0$) and with the Minkowski space-time metric $g = \begin{pmatrix}1 & 0\\ 0 & -1\end{pmatrix}$:

$$a^2 \to -1,\ \mu \to 0,\ \text{and}\ \alpha\beta = 1.\tag{19}$$

This suggests that the parameter $\mu$ could be taken proportional to $R/\Lambda_0$ raised to some power, where $\Lambda_0$ is a reference cosmological constant.

We consider here a special case of (17) which is a two-parameter model where $\gamma^1 = ia\begin{pmatrix}1 & 0\\ 0 & -1\end{pmatrix}$ and $\gamma^0 = \begin{pmatrix}i\mu a & \alpha\\ \alpha^{-1} & -i\mu a\end{pmatrix}$ resulting in the metric tensor $g = \begin{pmatrix}1-\mu^2 a^2 & -\mu a^2\\ -\mu a^2 & -a^2\end{pmatrix}$. Writing the two components of the spinor wave function in the Klein-Gordon equation (9) in terms of a single scalar function as $\psi_\pm(x,t) = e^{i\varepsilon(\mu x - t)}\phi(x)$, we obtain the following second order differential equation

$$\left[\frac{d^2}{dx^2} + \frac{1}{4a^2}\left(a'^2 - 2aa''\right) + \frac{\varepsilon^2 - m^2}{a^2}\right]\phi(x) = 0.\tag{20}$$

If we take $a(x) = e^{-\sqrt{R}x^2}$, which satisfies the flat space limit ($\lim_{R\to 0} a = 1$) then we obtain

$$\left[\frac{d^2}{dy^2} + \left(\frac{\varepsilon^2 - m^2}{R}\right)e^{2y} - \frac{1}{4}\right]\phi(x) = 0,\tag{21}$$

–6–

where $y = \sqrt{R}\,x$. In terms of the new variable $z = \frac{\zeta}{\sqrt{R}} e^y$, where $\zeta = \sqrt{\varepsilon^2 - m^2}$, this equation becomes

$$\left(z^2 \frac{d^2}{dz^2} + z \frac{d}{dz} + z^2 - \frac{1}{4}\right)\phi(z) = 0. \tag{22}$$

The solution is obtained directly as

$$\phi(z) = A\, J_{\frac{1}{2}}(z) + B\, Y_{\frac{1}{2}}(z) = \sqrt{\frac{2}{\pi z}}\left(A \sin z - B \cos z\right), \tag{23}$$

where $A$ and $B$ are normalization constants. $J_\nu(z)$ and $Y_\nu(z)$ are the Bessel functions of the first and second kind, respectively. Thus, for $\varepsilon^2 > m^2$ the variable $z$ is real and we obtain oscillatory solution with decreasing wavelength and decaying amplitude that goes like $e^{-\frac{1}{2}\sqrt{R}\,x}$ as the particle moves away from the origin. On the other hand, for $\varepsilon^2 < m^2$, $z$ becomes pure imaginary and the solution of Eq. (22) becomes a linear combination of $e^{\pm |z|}/\sqrt{z}$ of which only the non-diverging solution $e^{-|z|}/\sqrt{z}$ is allowed.

Now, the two-component Dirac equation (4) for the same two-parameter curved space model and with $\psi_\pm(x,t) = e^{i\varepsilon(\mu x - t)}\phi_\pm(x)$ gives

$$\left[\varepsilon \gamma^0 + i\gamma^1 \left(\frac{d}{dx} + i\mu\varepsilon\right) - \frac{i}{2}\frac{d\gamma^1}{dx} - m\right]\phi = 0. \tag{24}$$

Multiplying from left by $\gamma^1$ and using $(\gamma^1)^2 = g^{11} = -a^2$ and $\gamma_1' = \frac{a'}{a}\gamma_1$, we obtain the following matrix equation

$$\left[\frac{d}{dx} - \frac{a'}{2a} + \frac{m}{a}\begin{pmatrix}1 & 0 \\ 0 & -1\end{pmatrix} + \frac{\varepsilon}{a}\begin{pmatrix}0 & -\alpha \\ \alpha^{-1} & 0\end{pmatrix}\right]\begin{pmatrix}\phi_+ \\ \phi_-\end{pmatrix} = 0, \tag{25}$$

that results in the following relation between the two spinor components

$$\phi_\mp(x) = \pm \alpha^{\mp 1} \frac{a}{\varepsilon}\left(\frac{d}{dx} - \frac{a'}{2a} \pm \frac{m}{a}\right)\phi_\pm(x). \tag{26}$$

Using this relation to substitute one component in terms of the other in the two coupled first order differential equations resulting from (25) show that both components satisfy the following second order differential equation,

$$\left[\frac{d^2}{dx^2} + \frac{1}{4a^2}(a'^2 - 2a a'') + \frac{\varepsilon^2 - m^2}{a^2}\right]\phi_\pm(x) = 0, \tag{27}$$

which is identical to the Klein-Gordon equation (20). Therefore, taking $a(x) = e^{-\sqrt{R}\,x}$ results in an oscillatory solution of the form $e^{\pm iz}/\sqrt{z}$ for $\varepsilon^2 > m^2$ and a non-oscillatory solution of the form $e^{-|z|}/\sqrt{z}$ for $\varepsilon^2 < m^2$. Substituting these $\phi_\pm(x)$ in Eq. (26) gives $\phi_\mp(x)$ and results in two independent spaces of solutions for each energy range. For $\varepsilon^2 > m^2$, we obtain

(1) Positive energy: $\quad \psi(x,t) = \dfrac{e^{i\varepsilon(\mu x - t)}}{\sqrt{z}}\begin{pmatrix} A_+ e^{iz} + A_- e^{-iz} \\ \frac{m+i\zeta}{\alpha\varepsilon} A_+ e^{iz} + \frac{m-i\zeta}{\alpha\varepsilon} A_- e^{-iz} \end{pmatrix},$ (28a)

(2) Negative energy: $\quad \psi(x,t) = \dfrac{e^{i\varepsilon(\mu x - t)}}{\sqrt{z}}\begin{pmatrix} \alpha\frac{m-i\zeta}{\varepsilon} A_+ e^{iz} + \alpha\frac{m+i\zeta}{\varepsilon} A_- e^{-iz} \\ A_+ e^{iz} + A_- e^{-iz} \end{pmatrix},$ (28b)

where $A_\pm$ are normalization constants. On the other hand, for $\varepsilon^2 < m^2$, and with $\zeta = \sqrt{m^2 - \varepsilon^2}$ we obtain

–7–

(1) Positive energy: $\psi(x,t) = Ae^{i\varepsilon(\mu x-t)} \dfrac{e^{-z}}{\sqrt{z}} \begin{pmatrix} 1 \\ \frac{m-\zeta}{\alpha\varepsilon} \end{pmatrix}$, (29a)

(2) Negative energy: $\psi(x,t) = Ae^{i\varepsilon(\mu x-t)} \dfrac{e^{-z}}{\sqrt{z}} \begin{pmatrix} \alpha\frac{m+\zeta}{\varepsilon} \\ 1 \end{pmatrix}$. (29b)

It is worth noting that the overall factor $e^{i\varepsilon(\mu x-t)}$ in the wavefunction solutions of the Klein-Gordon and Dirac equations points to an effective phase velocity that is equals to $c/\mu$, where $c$ is the speed of light. Finally, an interesting special case of this two-parameter curved space model is when $a = (\sqrt{R}x+1)^{-2\xi}$, where $\xi$ is a real parameter and for which Eq. (20) and Eq. (27) become

$$\left[ \frac{d^2}{dy^2} - \frac{\xi(\xi+1)}{y^2} + \left( \frac{\varepsilon^2-m^2}{R} \right) y^{4\xi} \right] \phi(y) = 0, \tag{30}$$

where $y = \sqrt{R}x+1$. This equation is equivalent to the radial Schrödinger equation for power-law potential at "zero energy", which is known to have an exact solution for $\varepsilon^2 > m^2$ [33]. However, here the potential strength is true energy dependent. For $\xi = -1$, $a'^2 = 2aa''$ and the equation becomes an inverse quartic potential at "zero energy", which also has an exact solution [33].

## 4. CONCLUSION

We introduced a matrix algebra involving the Dirac gamma matrices defined on curved space and inspired by two notions: (1) unambiguous symmetric quantization leading to the Dirac equation in a curved space, and (2) squaring the Dirac equation results in the Klein-Gordon equation in its diagonal form and without first order derivatives. This scheme resulted in a specific and natural representation for the spin connections and allowed us to formulate the Dirac equation without the need for vierbeins. As an illustration of our findings, we studied the behavior of a relativistic particle in 1+1 space-time with a static metric. By choosing a special representation of the elements of the algebra, we obtained exact solutions for the Dirac and Klein-Gordon equations. The present work will shortly be extended and followed by a study of relativistic particles in 2+1 curved space in the presence of an electromagnetic field. An interesting application of such study is in connection with the work on massless Dirac fermions in graphene sheets in a magnetic field as well as in external potential structures.

## ACKNOWLEDGEMENT

We appreciate the generous support provided by the Saudi Center for Theoretical Physics (SCTP). We also acknowledge partial support by King Fahd University of Petroleum & Minerals (KFUPM).